# The scaling of the microLED and the advantage of 2D materials

Authors: Xuejun Xie[1], Hamid T. Chorsi[1], Kunjesh Agashiwala[1], Hsun-Ming Chang[1], Jiahao Kang[1], Jae Hwan Chu[1], Ibrahim Sarpkaya[2], Han Htoon[2], Jon A. Schuller[1], Kaustav Banerjee[1]*

[1]Department of Electrical and Computer Engineering, University of California, Santa Barbara, CA, USA

[2]Center for Integrated Nanotechnologies, Materials Physics and Applications Division, Los Alamos National Laboratory, Los Alamos, NM, USA

*Corresponding author (email: kaustav@ece.ucsb.edu)

**Abstract**

The demand for higher resolution displays drives the demand for smaller pixels. Displays show a trend of doubling the pixel number every 4 years and doubling the pixel per inch (PPI) every 6 years. As the prospective candidate for next-generation display technology, microLED (micro Light Emitting Diode) will suffer from sidewall current leakage and poor extraction efficiency as its lateral size reduces. Using Finite Element Analysis (FEA) method and Finite-Difference Time-Domain (FDTD) method, we find that reducing the thickness of the LED can reduce the current leaking to the sidewalls and reduce the total internal reflection simultaneously. A promising solution to this problem is by using atomically thin 2D materials to make LEDs. However, monolayer inorganic 2D materials that can provide red, green and blue emission are still lacking. Based on the blue light-emitting material fluorographene (CF), partially fluorinated graphene ($CF_x$) is synthesized in this work to emit red and green colors from 683 nm to 555 nm (limited by the instrument). This work also demonstrates lithographically defined regions with different colors, paving the way for the scaling of microLED.

Human vision consumes most of the energy of the human brain.[1] Displays are the channels that conduct information from the digital world to the human brain in the physical world. They drive the demand and innovation on content generation (image sensor), distribution (telecommunication bandwidth), and computing power (graphic processing unit, storage, I/O speed). Displays are becoming the driving force that pushes the semiconductor industry forward. Display modules are not only one of the most expensive components in smartphones,[2] but also consume the majority of battery power.[3] Displays have evolved from CRT (Cathode Ray Tube) to LCD (Liquid Crystal Display) to PDP (Plasma Display Panel) to AMOLED (Active-Matrix Organic Light-Emitting Diode) and eventually to the future microLED. With the revolution in technology from PC to laptop to smartphones, displays have gotten closer and closer to the human eyes. Moreover, displays have more pixels and smaller pixels over time. As summarized in **Figure 1a** and **1b**, like Moore's law of transistor scaling,[4] throughout the history of personal computing, the displays in the devices available in the market, such as personal computers, laptops, tablet pc, smartphones, and Virtual Reality (VR) headsets, show a trend of doubling the pixel number every 4 years and doubling the PPI (pixel per inch) every 6 years. (Some of the data are extracted from "pixensity.com"). The emergence of VR and Augmented reality (AR) as the next generation of personal mobile computing platform requests much more pixels, more PPI and faster refresh rate since they need to satisfy larger field of view in a much smaller form factor than traditional display,[5] as also discussed in **Supplementary Information S1**.

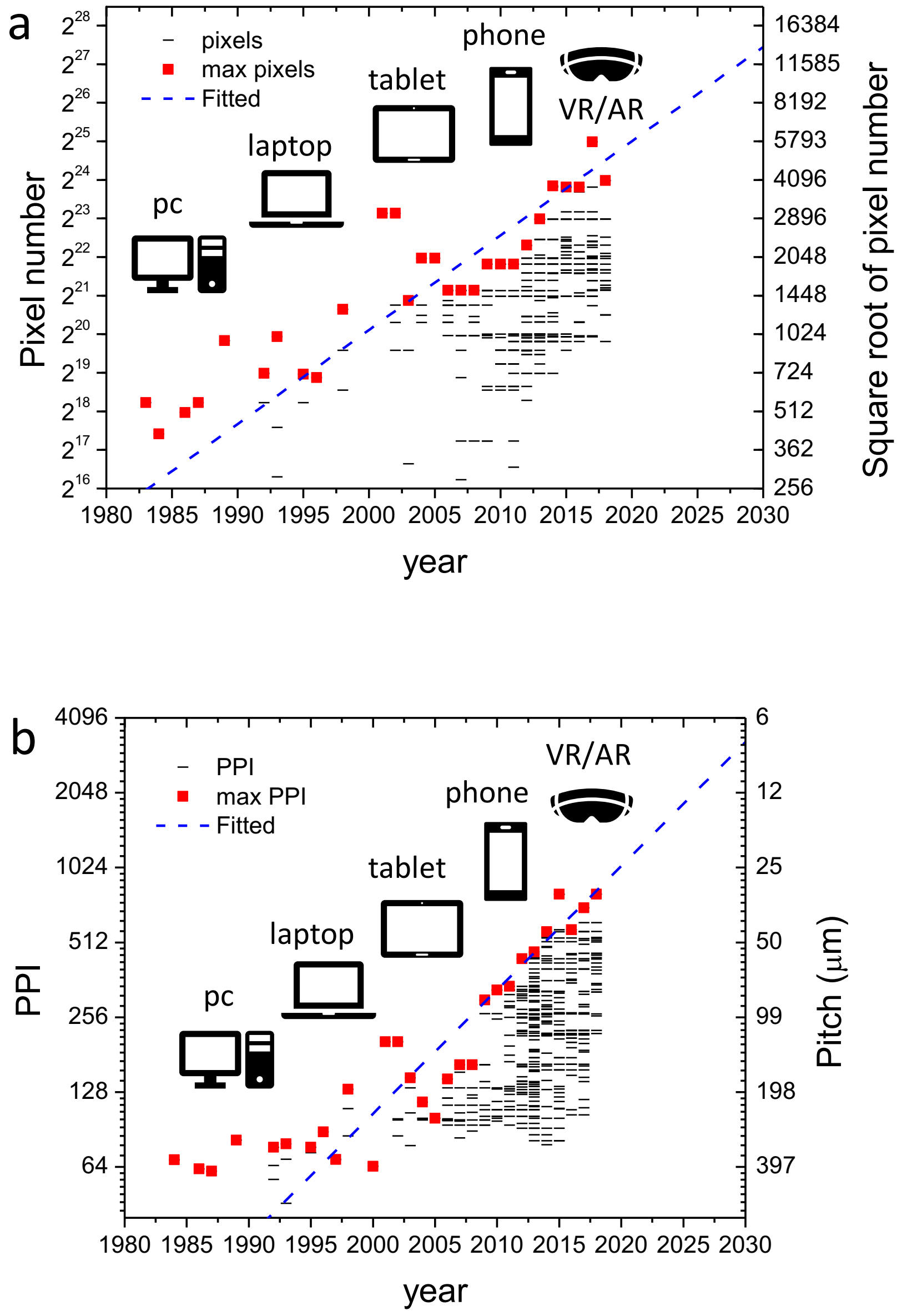

**Figure 1.** The summary of the total number of pixels and PPI for the consumer devices available in the market including TV, PC monitor, laptop, tablet, smartphone, and VR headset from 1983 to 2019. **(a)** The pixel number *vs*. year. The right axis is the square root of pixel number, which roughly indicates the number of pixel rows or columns. The fitted line is $pixels = 2^{year/4.09-468.62}$, in other words, the maximum pixel number doubles every 4.09 years. **(b)** The PPI *vs*. year. The right axis is the pitch of the pixel (the distance

between the centers of two adjacent pixels). The fitted line is $PPI = 2^{year/6.05-323.87}$, in other words, the maximum PPI doubles every 6.05 years.

The latest display technology is microLED. MicroLED uses inorganic LED, which offers high light-emitting efficiency, long lifetime, high refresh rate, and high pixel density.[6] In the microLED display, each pixel is made of red, green, and blue LEDs and controlled by transistors as shown in **Figure S2** in **supplementary information**. There are many ways to create MicroLED displays, either by transferring LED dies on top of control circuit which could be on glass substrate or silicon substrate, or by creating control circuit on top of LED wafer. Different processes have different challenges to scale down microLED display.[7] For example, for transferring technique to glass substrate, the accuracy of transfer process limits the scaling. For transferring to silicon substrate, the bonding process limits the scaling. For creating control circuit on top of LED wafer, the capability of reducing the gate length of thin-film transistor limits the scaling. Regardless of the process issues, from the device physics point of view, the scaling down of the microLED pixel size requires scaling down of both its electrical and optical components. As discussed in **supplementary information S2**, the scaling down of electrical components (transistors) significantly improves display performance (PPI, refresh rate, brightness). However, the scaling down of the optical components (LED) has major two issues: The first issue is the sidewall defects-induced non-radiative recombination.[8], [9] In simpler words, as the LED scales smaller, the leakage current through the sidewall defects will not only reduce the electron injection efficiency but also lead to significant heat generation. The second issue is light-extraction efficiency. LEDs are usually made of high-refractive-index materials. Due to the total internal reflection, most of the light emitted by the active region gets trapped inside the semiconductor and turns into heat. This forms the fundamental physical limit on LED light-extraction efficiency. Those two issues can be solved simultaneously by reducing the thickness of the LED.

First, in order to analyze how the reduction in thickness reduces the sidewall leakage, let’s assume a homogeneous conductor with a rectangular cross-section as shown in the inset of **Figure 2a**. (The result is independent of the resistivity of the conductor and only depends on the geometry.) The current input $I_{in}$ is from the top face. The bottom face is the active region. The current leaks to the sidewalls as $I_{leak}$, and reaches the bottom edge as $I_{out}$. Assuming all $I_{in}$ are converted to photon, the electron injection efficiency is $I_{out}/I_{in}$. The simulation is carried out with FEA as detailed in **supplementary information S3**. As shown in **Figure 2a**, if the thickness of LED decreases, while assuming the sidewalls to be perfect conductors (worst case), the electron injection efficiency can still approach ~100% when the w/h is larger than 100 (for example, $w = 1\ \mu m$ and $h < 10\ nm$), which is the case for monolayer 2D materials whose thicknesses are around ~1 nm.

Second, to analyze the light-extraction efficiency, let us treat a light emitter as a slab waveguide made with high reflective index materials. It has been shown by FDTD simulation that if the LED thickness is reduced from 7.12 $\mu m$ to 1.12 $\mu m$, the light extraction efficiency can increase from 9.36% to 24.1%.[10] In this work, we calculated the light extraction efficiency when the thickness of LED is reduced to the nanometer scale, and the details about the calculations are shown in **supplementary information S4**. We simulated a dielectric slab (n = 2.4 the reflective index of GaN at 550 nm) with an embedded dipole source, where the vacuum wavelength $\lambda = 550\ nm$ ($n$ = 1). The simulation setup consists of a 20 $\mu m$ by 20 $\mu m$ slab, where the 4 sides are terminated with perfectly matched layers, and the thickness is varied. The dipole source is placed at the center of the dielectric slab and the thickness is swept from 1 nm to 180 nm. Figure S4 **supplementary information S4** shows the far-field angular radiation distribution for x oriented dipole. For thinner dielectric slab, it shows more radiation at larger angles. To account for a real-life situation, three dipole orientations (x/y/z) are considered and the average is recorded. As shown in **Figure 2b**, the results show that the light extraction efficiency is highly thickness dependent and one can achieve significant enhancement using thinner materials. A light extraction efficiency close to 100% can be achieved using atomically thin materials. The physics behind this is that as dielectric slab becomes thinner the slab waveguide modes are reduced. (A detail illustration is shown in **Figure S6** of **supplementary information S4**.) For the slab thinner than the single-mode limit, when the slab's thickness is future reduced, the mode overlap between the slab and the dipole radiation is future reduced. So thinner dielectric slab leads to less light getting trapped inside the dielectric slab. Moreover, from the device fabrication point of view, it's easier to pattern and etch small pixels with thin materials as compared to thick materials.

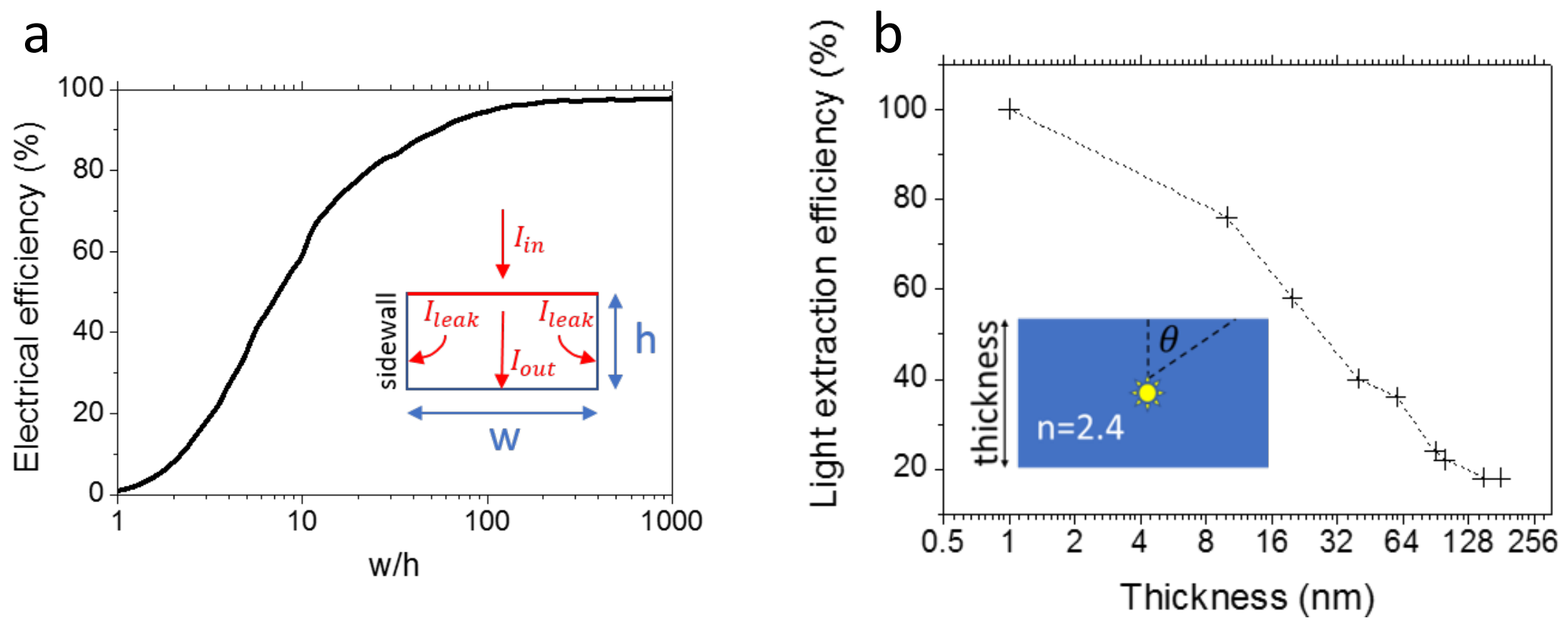

**Figure 2.** Solving the LED scaling issues by reducing the thickness of LED. **(a)** Electron injection efficiency due to the sidewall leakage. As the aspect ratio w/h increases, the Electron injection efficiency approaches 100% even the sidewalls are perfect conductors. **(b)** Light extraction efficiency as a function of thickness. The light emits from a dipole at the center of a thin slab with n = 2.4 and $\lambda$ = 550 nm. As the thickness reduces, the Light extraction efficiency approaches 100%. The details about the calculation are shown in **supplementary information S4**.

LEDs based on atomically thin 2D heterostructure with few nanometers of total thickness are the ultimate solution for thin pixels.[11] 2D materials have also been reported to have near-unity photoluminescence quantum yield, probably due to the lack of dangling bonds and low surface defects.[12] LEDs made with 2D materials are shown to have ultra-low threshold current density, due to the geometrically confined carriers.[13] Due to the atomically thin thickness, it's easier and more precise to create quantum dots on 2D materials for single-photon light-emitting device.[14], [15] However, it's still difficult to fabricate the monolayer 2D materials that can cover the entire visible light spectrum including red, green and blue colors.

As shown in **Figure 3a,** only a few inorganic monolayer 2D materials have bandgaps within the visible light range. Additionally, there are very few direct bandgap monolayer 2D materials, and they don't cover the red, blue, green region of the visible spectrum. One such 2D material of interest is fluorographene, a UV light-emitting direct bandgap semiconductor with 3.8 eV bandgap or 326 nm emission.[16] Many theoretical works predict that by tuning the size of graphene quantum dots inside fluorographene, the bandgap can be tuned from 0 to 3.8 eV.[17], [18] Several works attempted to use powder or solution process to create fluorographene with a tunable bandgap.[16], [19], [20] However, it's difficult to create a uniform light emitter from powder or solution-processed graphene, due to the random size of graphene flakes. Fluoridation of monolayer CVD (Chemical Vapor Deposition) graphene has been shown by using $XeF_2$ treatment.[21], [22] But no previous work looks into tuning the photoluminescence properties of fluorinated graphene, especially partially fluorinated graphene. In this work, by tuning the amount of the fluorine atoms on partially fluorinated monolayer graphene, green and red light-emitting materials (555 nm to 683 nm limited by instrument) are created. Additionally, using E-beam lithography, different regions of graphene can be patterned to possess tunable bandgaps due to different $XeF_2$ exposure time.

The experimental procedure can be summarized as follows. Monolayer graphene is synthesized on copper foil with CVD and then transferred to $SiO_2$ substrate. For detailed clarity, the transfer steps have also been elucidated here. First, one face of the copper foil with graphene is covered with PMMA950A4 by spin coating, and the other face is etched with oxygen plasma. Then the copper foil is floated on $(NH_4)_2S_2O_8$ water solution to remove the copper. Eventually, the graphene with PMMA is transferred onto 285 nm $SiO_2$/Si substrate and cleaned with acetone. The fluoridation of graphene is achieved using $XeF_2$ (Xetch-X3) with 60 seconds of treatment with various $XeF_2$ pressures varying from 0.1 T (Torr) to 4 T and repeated with different treatment cycle.[21], [22] **Figure 3b** shows the fluorescent microscope image of $CF_x$ treated with 60s of $XeF_2$ at 1T, where the $CF_x$ shows orange fluorescence with blue light excitation.

In order to control the light emission wavelength from $CF_x$, the following steps are performed. Firstly, the treatment repeating cycle is changed. As shown in **Figure 3c**, the PL peak wavelength does not change as the cycle number increases. Secondly, the treatment pressure is changed. **Figure 3d** shows the measured PL peak wavelength *vs*. $XeF_2$ pressure. The PL peak wavelength decreases as the treatment pressure increases. Because $XeF_2$ has a vapor pressure of ~4T at room temperature (higher pressure at room temperature will turn $XeF_2$ into liquid), the upper limit of the Xetch-X3 for the $XeF_2$ pressure is approximately 4T. Further increasing the pressure at higher temperature could help in further increasing the bandgap and reducing the PL peak wavelength. Interestingly, the samples also show redshift after 15 days of aging in the air. The PL spectra from various samples are shown in **Figure 3e**. The spectra show light emission with peaks varying from 555 nm to 683 nm. (Note that, the red light emission spectra are extracted from 15 days old samples.) **Figure 3f** shows some of the Raman spectra of the various samples with different treatment pressure. As the $XeF_2$ treatment pressure increases, the FWHM (full width at half maximum) of D peaks increases as shown in **Figure 3g**, which indicates graphene is more fluorinated at higher treatment pressure.[23] **Figure 3h** and **i** show the X-ray photoelectron spectroscopy (XPS) signal of Fluorine and Carbon atom, respectively. As the $XeF_2$ treatment pressure increases, the area of F peak increases and that of C-F peak increases comparing to the C-C peak, which again implies that more F atoms are being attached to C atoms. The F atom ratio "x" in $CF_x$ is extracted from XPS measurement by calculating the area of the peaks as shown in **Figure 3j**. The F atom concentration increases as the $XeF_2$ treatment pressure increases. After 15 days of aging, the sample shows less "x", which means fewer F atoms. Although the underlying mechanism of redshift over time is not very clear so far, this phenomenon could potentially be used to tune the color of $CF_x$ and a better encapsulation method could help stabilizing $CF_x$. As summarized in **Figure 3a**, this work covers the spectrum from red to green and has potential to create a blue color emitter.

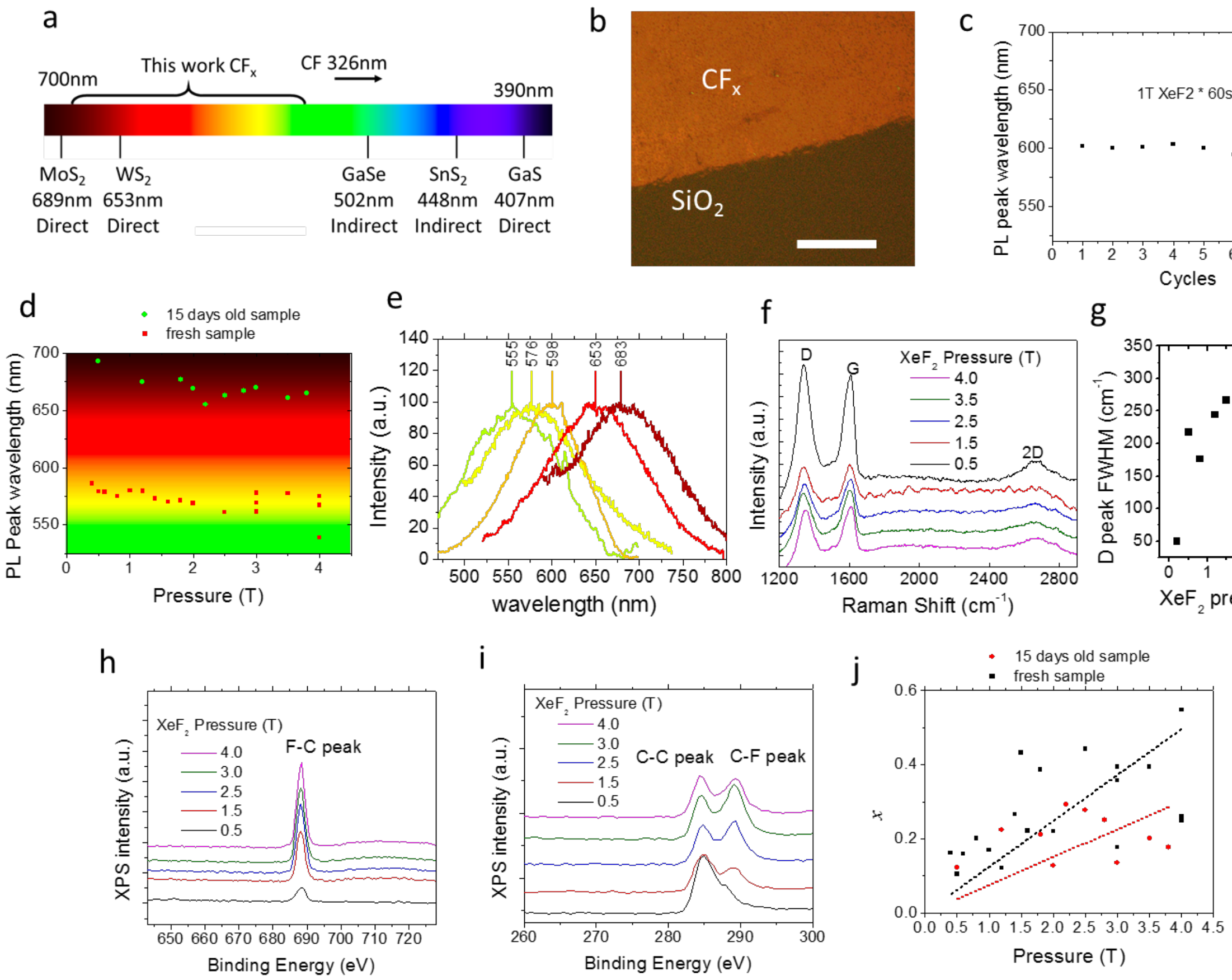

**Figure 3. (a)** Bandgaps of various monolayer 2D materials in the visible light range. Only a few inorganic monolayer 2D materials have the bandgap within visible range. This work covers the red to green spectral range and has the potential to further extend to blue. **(b)** Fluorescent microscope image of partially fluorinated graphene ($CF_x$) with 60s 1T $XeF_2$. The $CF_x$ shows an orange color. The white scale bar is 100 μm. **(c)** The peak wavelength *vs*. the $XeF_2$ treatment cycle. The PL peak doesn't have strong correspondence on the treatment cycle. **(d)** The peak wavelength *vs*. $XeF_2$ treatment pressure. As the pressure increases, the PL peak shows blueshift. It's interesting to find that, after 15 days of aging in air, the samples show redshift. **(e)** Photoluminescence spectra of $CF_x$ with emission peaks from 555 nm to 683 nm. **(f)** The Raman spectrum with different treatment pressure. **(g)** FWHM of the D peak in (f). As the $XeF_2$ treatment pressure increase, the D peak becomes wider. **(h)** The XPS signal of Fluorine atom. **(i)** The XPS signal of carbon atom. The number of fluorinated carbon (C-F) bond increases as the $XeF_2$ treatment pressure increases. **(j)** F atom

ratio of the total atom "$x$" as a function of $XeF_2$ pressure extracted from XPS measurement. After 15 days, the $x$ in the samples reduces.

To make pixel arrays for display, it's very important to be able to define different colors in different regions. This can be achieved using E-beam lithography, the details of which are as  as follows: First, the photoresist (PMMA950A4) is applied to monolayer graphene on $SiO_2$/Si substrate by 3000 rpm spin coating for 30s followed by 180°C baking for 60s. Then the it is exposed to e-beam lithography with 350 μC/cm$^2$ area dose and developed with isopropanol (IPA) and methyl isobutyl ketone (MIBK) (IPA:MIBK = 1:3). The result is shown in **Figure 4a**. Second, the patterned sample is treated with 4T $XeF_2$ for 60s where the black regions in **Figure 4a** are exposed. The PMMA in the bright region in **Figure 4a** is later washed with acetone. The result is shown in **Figure 4b**. The fluorinated region has larger bandgap, thus lesser absorption, and more transparency, thus showing up brighter under microscope. Third, the sample is treated with 1T $XeF_2$ for 60s. As shown in **Figure 4c**, the pattern becomes undistinguishable under visible light. However, under blue light fluorescent microscope, different regions with different $XeF_2$ treatment pressure become distinguishable. As shown in **Figure 4d**, the "UCSB" logo shows orange color while the surrounding region shows green color. In Kelvin probe force microscopy (KPFM) measurement shown in **Figure 4e** and **4f**, two different regions show different contact potential difference ($V_{CPD}$). The region with larger $XeF_2$ pressure treatment shows lower $V_{CPD}$, in other words, lower work function. The reason is that higher $XeF_2$ pressure creates larger bandgap, which makes the workfunction is higher.

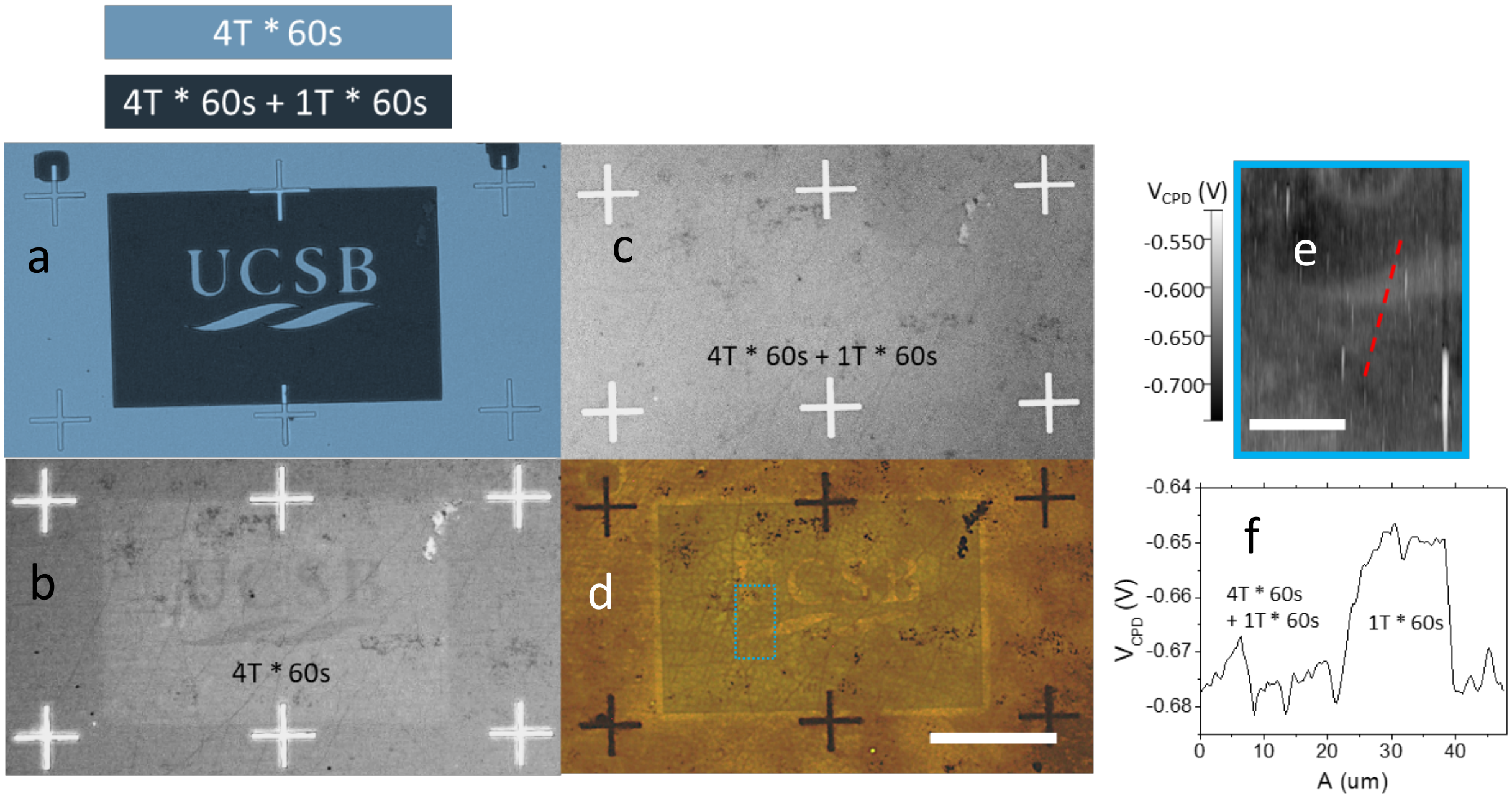

**Figure 4. (a)** The optical microscope image of a designed pattern after E-beam lithography. The dark regions are exposed regions. The color bars denote the different regions with different treamtment conditions. **(b)** The optical microscope image of the patterned sample treated with 4T $XeF_2$ for 60s and washed with acetone under microscope with halogen light illumination. The treated region is brighter due to less light absorption of $CF_x$ with larger bandgap comparing to graphene. **(c)** The optical microscope image of the sample treated with another 1T $XeF_2$ for 60s. The pattern is not distinguishable under microscope with halogen light illumination. **(d)** Fluorescent microscope image of patterned graphene with blue light illumination (Excitation: 450-480nm / Dichroic 500nm / Barrier 515nm+). The "UCSB" logo lights up in orange color, while the surrounding region shows green color. **(e)** KPFM measurement of the patterned region as marked with blue dash line in (d). **(f)** Cross-sectional plot of KPFM measurement marked with red dashed line in (e). The region with higher $XeF_2$ treatment pressure shows lower workfunction.

As a summary, the trend, the challenge, and the prospects of microLED scaling are revealed for the first time. For the consumer electronic devices in the market including personal computers, laptops, tablet PC, smartphones, and Virtual Reality (VR) headsets, the pixel number doubles every 4 years and the PPI doubles every 6 years. As the microLED leads the innovation of next-generation display, the challenge is making smaller LEDs. By reducing the thickness of LED, not only can it overcome edge defect induced leakage but also increases the out-coupling efficiency. The ultimate solution is using 2D materials for atomically thin light emitter. The first challenge is creating 2D materials with desired bandgap to be able to emit light in visible spectrum range. The method to control the PL peak wavelength by $XeF_2$ treatment on graphene is studied. Our experiments reveal that it is possible to tune the light emission of $CF_x$ from 555 nm to 683 nm with the potential to cover the entire visible light spectrum range. This method can also work with lithography techniques to define multi-color emitters in different regions with different $XeF_2$ exposure pressure. This work paves the way for next wave of innovation on the scaling of pixels.

**Acknowledgment**

This research was supported by the UC Lab Fees Research Program (grant LFR-17-477237). The DFT simulations were performed in the Center for Scientific Computing at UCSB that is supported by the NSF Grant CNS-0960316.

**Supplementary information for: The scaling of microLED, the challenges and the prospect**

Authors: Xuejun Xie[1], Hamid T. Chorsi[1], Kunjesh Agashiwala[1], Hsun-Ming Chang[1], Jiahao Kang[1], Jae Hwan Chu[1], Ibrahim Sarpkaya[2], Han Htoon[2], Jon A. Schuller[1], Kaustav Banerjee[1]*

[1]Department of Electrical and Computer Engineering, University of California, Santa Barbara, CA, USA

[2]Center for Integrated Nanotechnologies, Materials Physics and Applications Division, Los Alamos National Laboratory, Los Alamos, NM, USA

*Corresponding author (email: kaustav@ece.ucsb.edu)

**S1. The requirement of the pixel scaling.**

To make a compact display that fit inside a headset with a small form factor, it's critical to make the pixel as small as possible. The lower limit of pixel pitch is determined by the diffraction limit of the numerical aperture (NA) of the optical system as shown in **Figure S1a**. For example, for a microlens with NA=0.99,[1] the optical resolution for green color ($\lambda$ = 550 nm) is determined by $\lambda/2NA$, which is 275 nm or 92364 PPI. So the pixel pitch needed for future light-field display need to be much smaller than the current technology.[2] On the other hand, the display for augmented reality (AR) brings new challenges on display's refresh rate. For AR headset, since the display is moving along with the user's head with real-world as background reference, the images on the display are constantly moving when user moves. The human eye can trace objects with angular speed of 100°/s respect to the eye.[3] If the display dosen't catch up with this speed, ghosting effect will appear, where image has a mismatch with the real world. For distinguishing a ghosting with 5 milliradians or 0.286 degrees, the delay (or latency) between physical movement to the digital image refreshing should be smaller than $\frac{0.286^o}{100^o/s} = 2.86$ ms or 350 Hz.[4] Human eyes have the capability to distinguish 0.3 milliradians or 1/60 degree, as illustrated in **Figure S1b**. To provide undistinguishable image without any delay, the maximum limit of refresh rate is $\frac{100^o/s}{1/60^o} = 6000$ Hz.

The required PPI and refresh rate of display for an AR application is summarized in **Figure S1c**

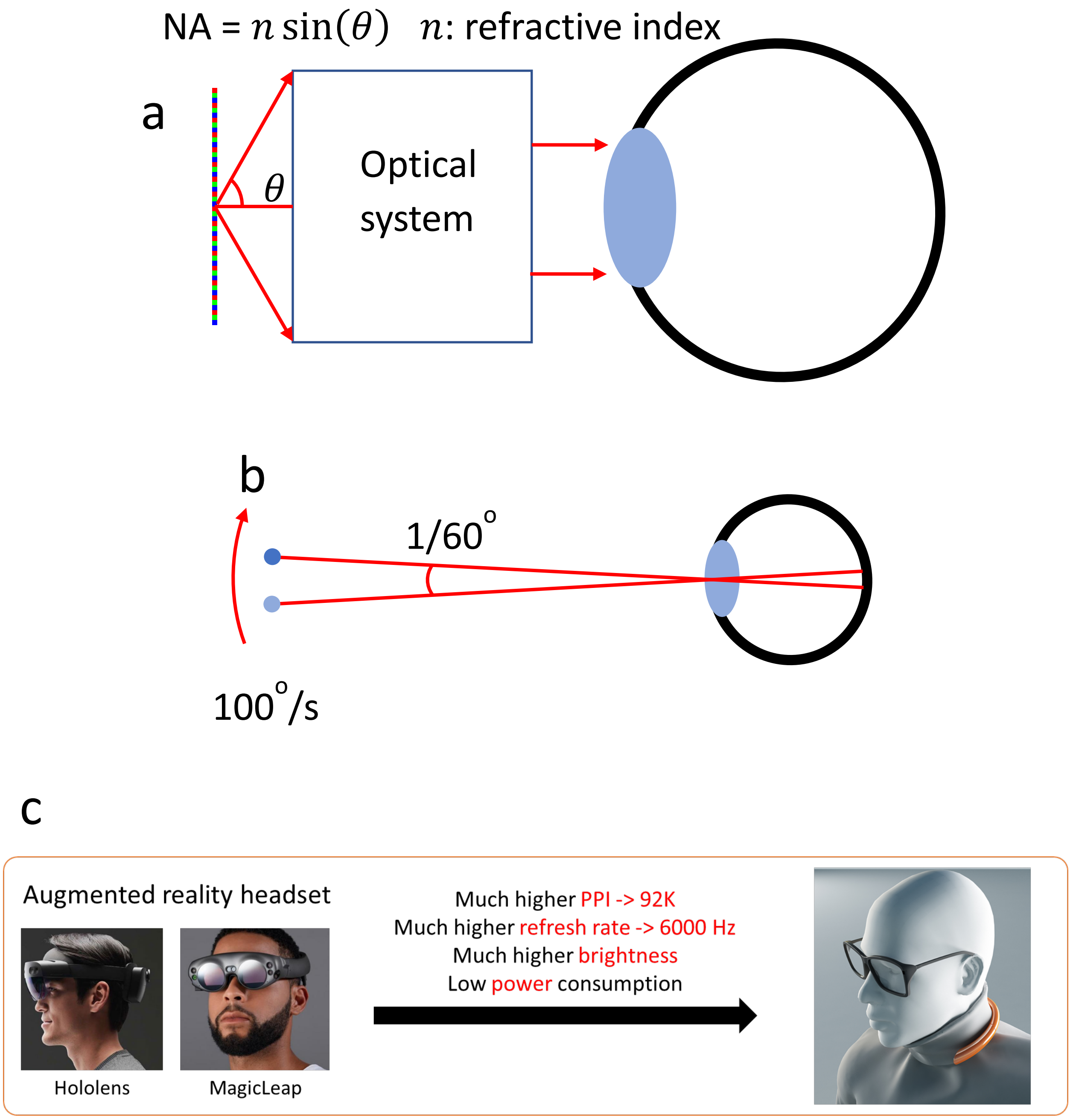

**Figure S1. (a)** The typical AR optical system. The required minimum pixel size is determined by the NA of the optical system. **(b)** The illustration of the up limit of the speed of moving an object for human eyes. **(c)** PPI and refresh rate of display for AR applications.

## S2. Scaling of the pixel circuit

The display industry usually depends on 1:1 projection lithography to create a large panel. But in the age of augmented reality, microdisplay is key components. Microdisplay uses stepper for lithography, similar to the VLSI technology. Like the scaling of VLSI, microdisplay can also benefit from the reduction of minimum line width as shown in **Figure S2**. For example, 2T-1C (2 transistors and 1 capacitor) pixel circuit is the basic circuit for driving self-emitting pixels like OLED (organic light-emitting diode) and microLED (micro light-emitting diode). The scaling of 2T1C circuit is illustrated in Figure 1c. If the pixels scale down by 0.5X, the driving power of T1 doesn't change. That means the maximum current that can be provided by the transistor does not change. But the LED's size has been reduced by 1/4X. Thus, the current density enhances by 4X, implying that the scaled LED can be 4 times brighter assuming the LED brightness is linearly proportional to the driving current. On the other hand, the capacitance C becomes 1/4X, so the pixel can be refreshed 4 times faster. Since the pixel row number is doubled, the frame rate of the display can also be 2 times faster.

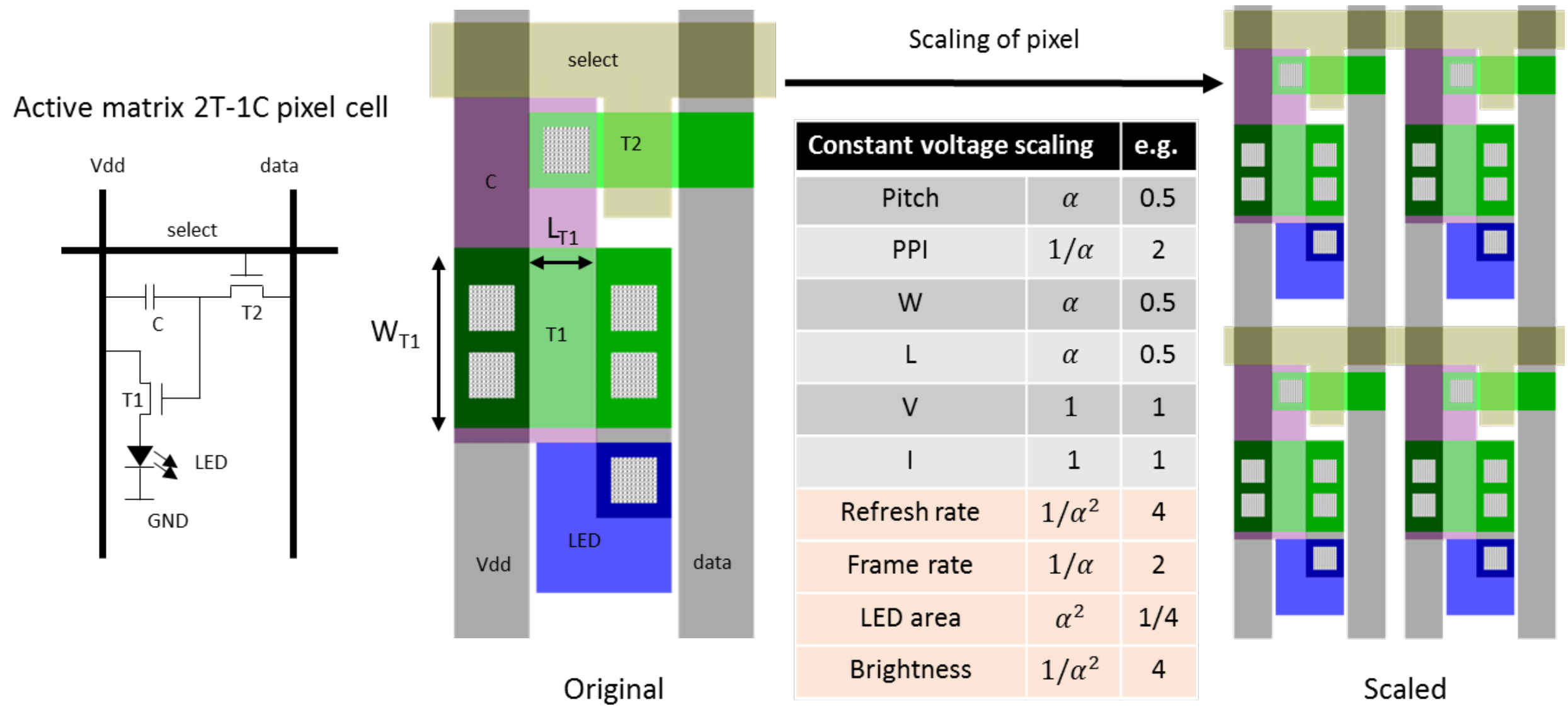

| Constant voltage scaling | | e.g. |
|---|---|---|
| Pitch | $\alpha$ | 0.5 |
| PPI | $1/\alpha$ | 2 |
| W | $\alpha$ | 0.5 |
| L | $\alpha$ | 0.5 |
| V | 1 | 1 |
| I | 1 | 1 |
| Refresh rate | $1/\alpha^2$ | 4 |
| Frame rate | $1/\alpha$ | 2 |
| LED area | $\alpha^2$ | 1/4 |
| Brightness | $1/\alpha^2$ | 4 |

**Figure S2.** The left-hand side illustrates the typical active matrix circuit with 2T-1C design. T1 is used to drive the LED and T2 with the capacitor are used to control T1. The right-hand side is the illustration of layout scaling of 2T-1C circuit. As the layout scales down, the improvement of the performance is summarized in the table.

**S3. Comparing thick and thin sample on sidewall leakage**

Here we assume the extreme case, where the sidewall becomes a perfect conductor. We solve a 3D-PDE with Laplace equation ($\nabla^2 \cdot u = 0$ where $u$ is the electric field) and Dirichlet boundary condition by FEA calculation assuming the side and bottom edge are perfect grounded conductor. As shown in **Figure S3**, the thick sample shows great amount of current reaching the edge. While the thin sample shows suppressed edge leakage.

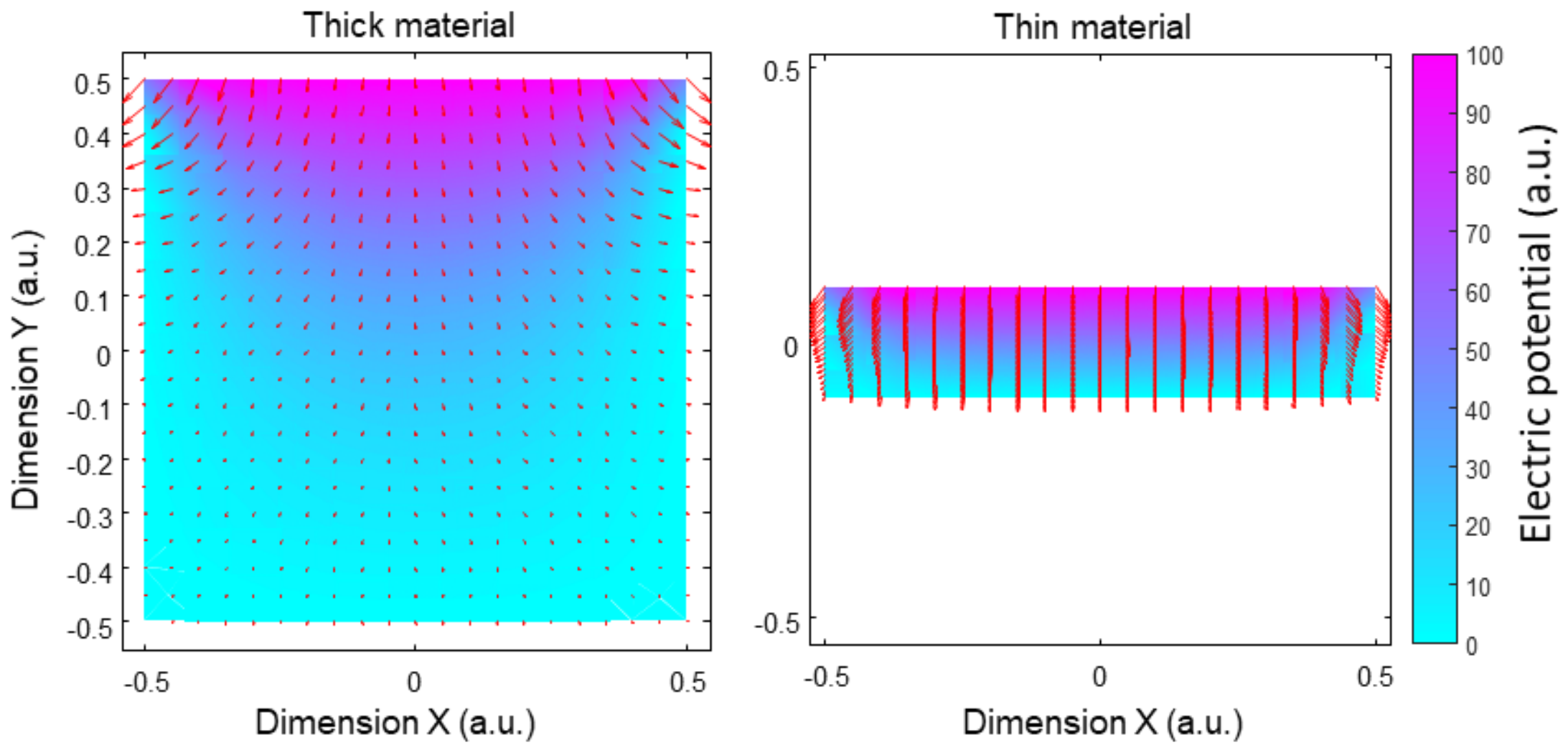

**Figure S3.** Cross-section of thick and thin LED. The color scale is the percentage of voltage. The red arrows show the current density. The top face is biased, and the side and the bottom are grounded. The thin device shows less current leaking to the sidewall. (All the scales are relative number.)

**S4. Simulation of light extraction from a thin layer of dielectric.**

Finite-difference time-domain method (FDTD) based simulations of extraction efficiency were performed by Lumerical with the structure as shown in **Figure S4a**. The electric dipole (ED) source was placed at the center of the dielectric slab with a refractive index of 2.4 (which is the reflective index of GaN at 550 nm) and the thickness was swept from 180nm down to 1nm. Air was used as the background. Perfectly matched layer (PML, absorbing) boundaries were used in all four boundaries for our 2D simulations. Dipole sources with x, y, z, orientations were considered, and the average of emitted power was recorded. The simulation results for different thickness and different wavelength is shown in **Figure S4b** and **S4c**.

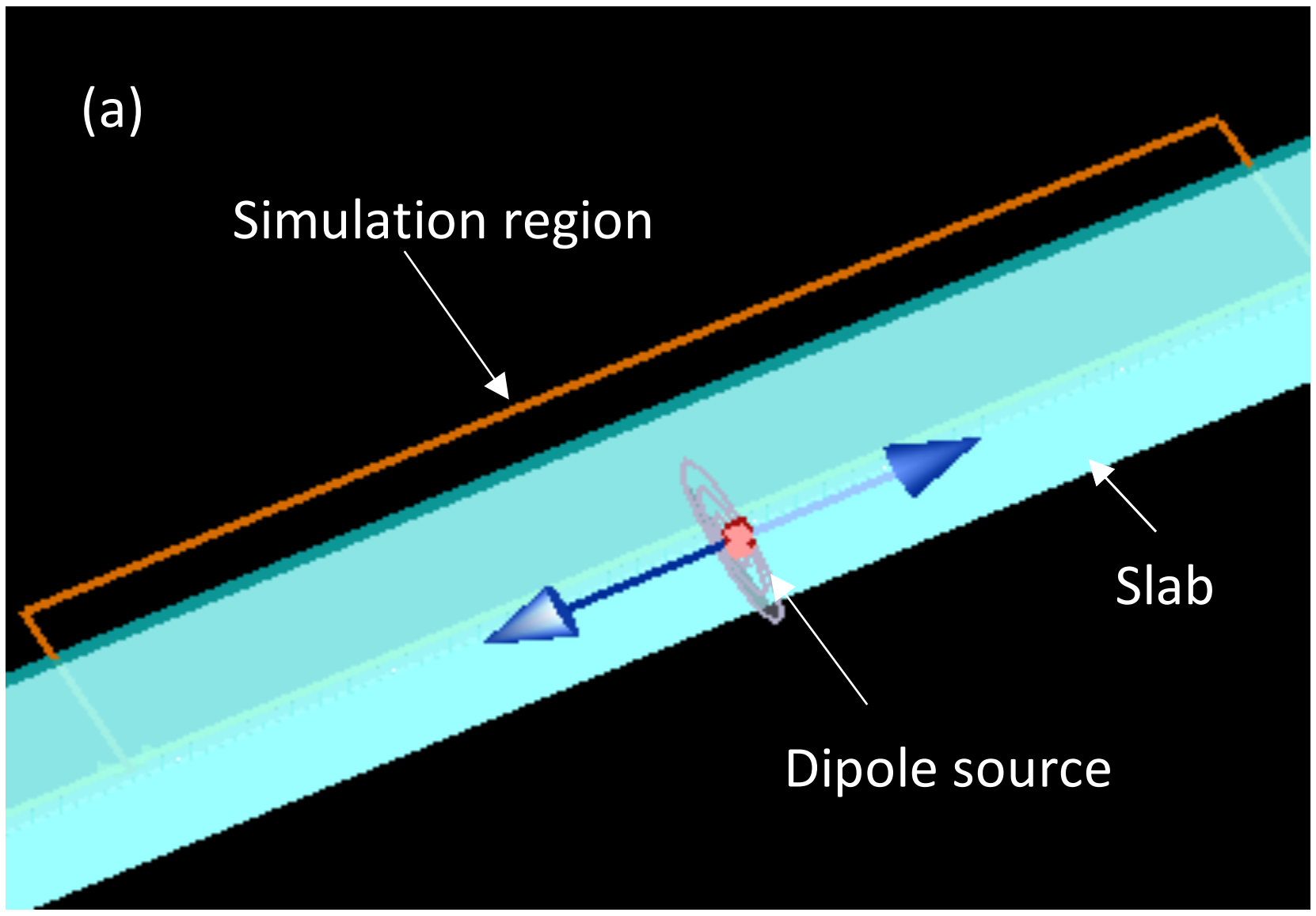

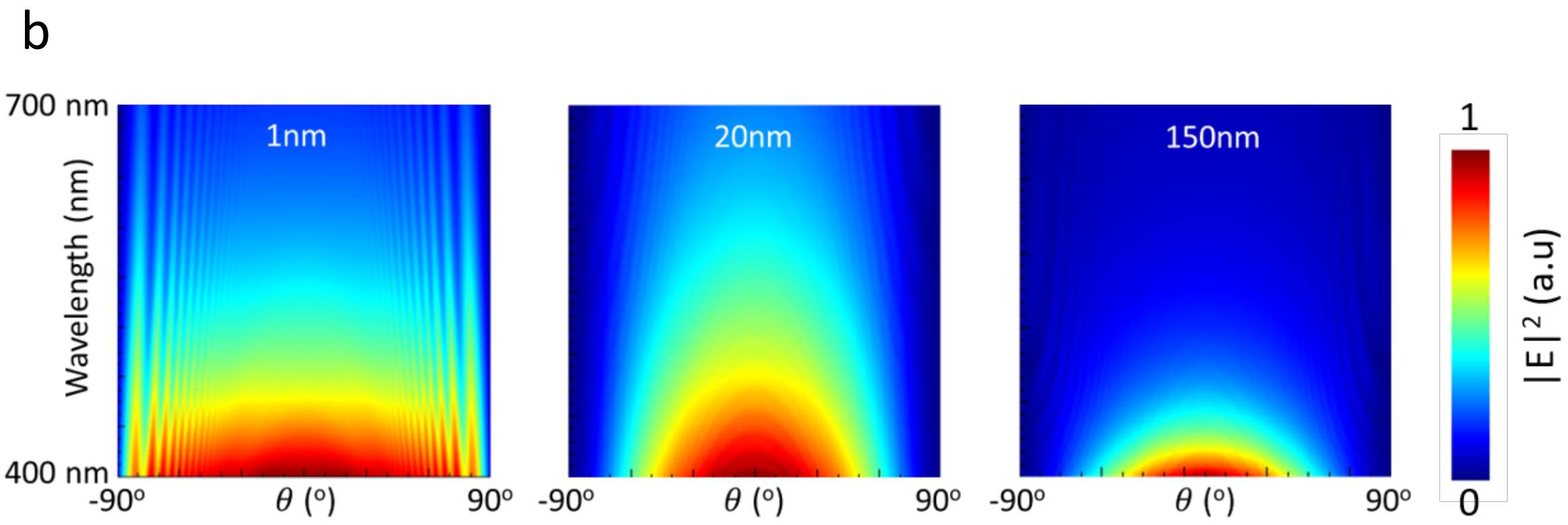

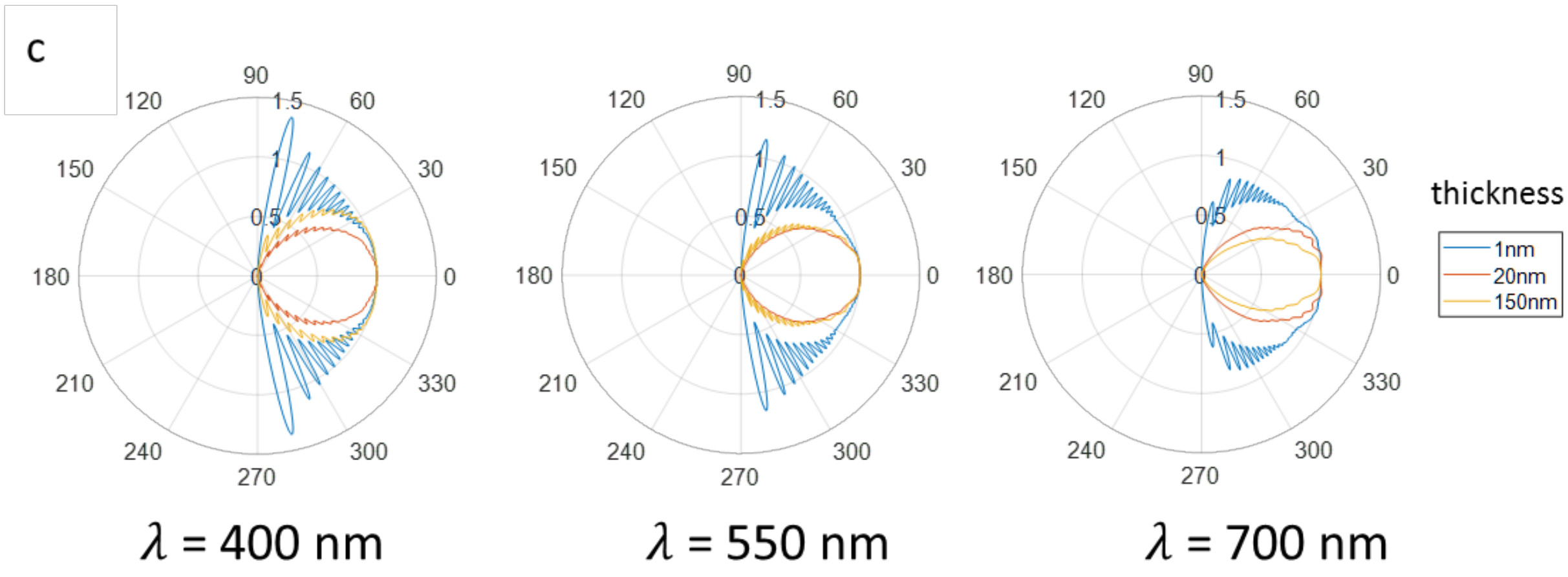

**Figure S4**. **(a)** Schematic of simulation region. **(b)** and **(c)** shows far-field angular distribution at a different wavelength. It's clear to see that in thinner LED light with larger $\theta$ can escape.

The light extraction efficiency was obtained by the ratio of the total external power emitted (light escaping to the air) to the emitted power from the dipole.

$$\eta = \frac{P_{external}}{P_{source}}$$

The light extraction efficiency is depending on many parameters including the optical properties of the material in the active and background regions, number of layers, roughness of the interface, wavelength, and the thickness of the active layer. **Figure S5** shows the wavelength dependent on extraction efficiency for a material with a refractive index of 2.4 at different thicknesses. As mentioned in the main manuscript, the extraction efficiency increases by decreasing the thickness of the active layer.

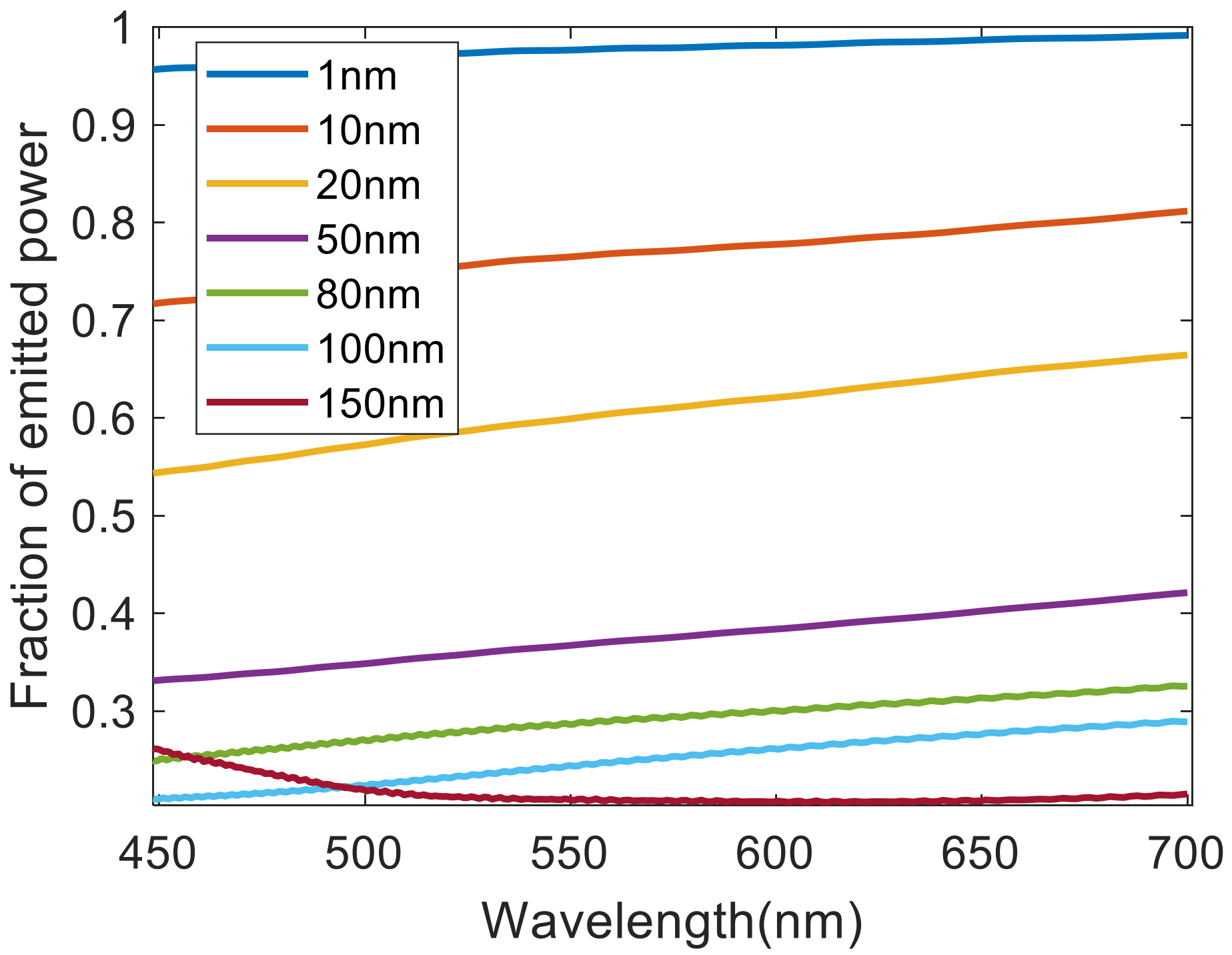

**Figure S5**. Wavelength dependent and thickness dependent light extraction efficiency.

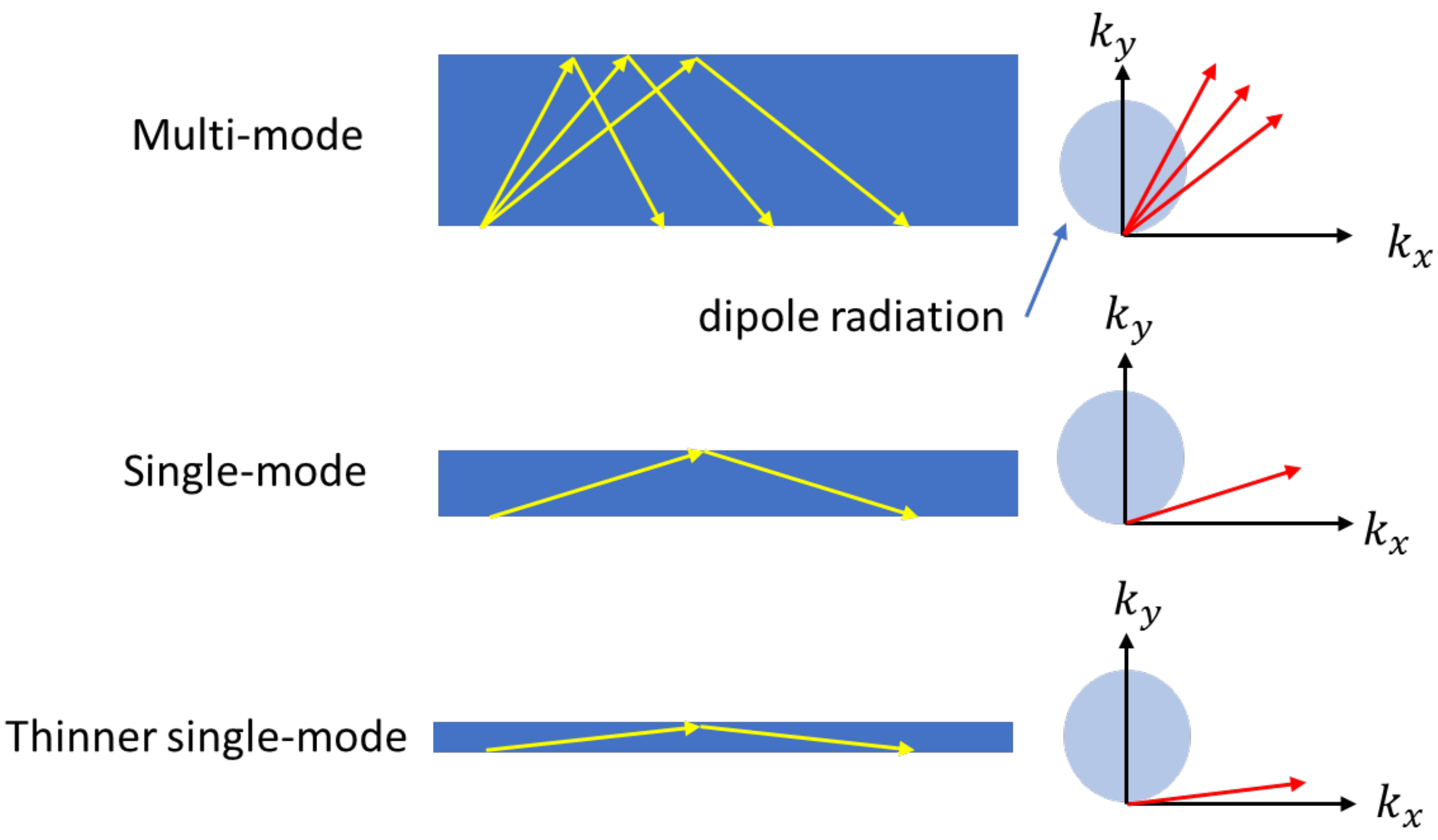

**Figure S6.** This drawing illustrates the physics behind the FDTD simulation. The mode overlaps between dipole radiation (blue circle) and the guided modes (red arrows) inside the dialectic slab define how much energy is trapped inside the slab. Light with no mode overlap with the slab will escape from the slab. The smaller the mode overlap, the larger the extraction efficiency. As the dielectric slab becomes thinner, the slab changes from multi-mode to single-mode. The mode numbers reduce which reduces the mode overlap. As the single-mode getting thinner, the momentum of the guided mode rotates towards x axis, which further reduces the mode overlap with the dipole radiation.